\documentclass[a4paper,prl,twocolumn,showpacs,superscriptaddress,floatfix,nofootinbib]{revtex4-1}
\usepackage{lipsum}
\usepackage{graphicx}
\usepackage{epsf}
\usepackage{epsfig}
\usepackage{amssymb,amsmath,amsfonts}
\usepackage[usenames]{color}
\usepackage{amssymb}
\usepackage{times}
\usepackage{bm}
\usepackage{hyperref}
\hypersetup{
  colorlinks=true,        
  linkcolor=blue,         
  citecolor=cyan,         
}

\newcommand{\ITP}{Institut f{\"u}r Theoretische Physik,
  Max-von-Laue-Stra{\ss}e 1, 60438 Frankfurt, Germany}
\newcommand{\FIAS}{Frankfurt Institute for Advanced Studies,
  Ruth-Moufang-Stra{\ss}e 1, 60438 Frankfurt, Germany}
\newcommand{\KCCT}{Kobe City College of Technology, 651-2194 Kobe, Japan}
\newcommand{\IUCAA}{Inter-University Centre for Astronomy and
  Astrophysics, Post Bag 4, Ganeshkhind, Pune 411 007, India}
\newcommand{\WSU}{Department of Physics \& Astronomy, Washington State
  University, 1245 Webster, Pullman, WA 99164-2814, U.S.A \\}
\newcommand{\CARDIFF}{School of Physics and Astronomy, Cardiff
  University, 5, The Parade, Cardiff, UK, CF24 3AA}
\newcommand{\PSUPHYS}{Institute for Gravitation and Cosmos, Physics
  Department, Pennsylvania State University, University Park, PA, 16802,
  USA}
\newcommand{\PSUASTRO}{Department of Astronomy \& Astrophysics,
  Pennsylvania State University, University Park, PA, 16802, USA}
\definecolor{orange}{rgb}{1,0.5,0}

\newcommand{\cf}{cf.~}
\newcommand{\ie}{i.e.,~}
\newcommand{\eg}{e.g.,~}
\renewcommand{\BibitemShut}[1]{}
\begin{document}

\title{Neutron-star Radius from a Population of Binary Neutron Star Mergers}
\author{Sukanta~Bose}
\affiliation{\IUCAA}
\affiliation{\WSU}
\author{Kabir~Chakravarti}
\affiliation{\IUCAA}
\author{Luciano~Rezzolla}
\affiliation{\ITP}
\affiliation{\FIAS}
\author{B.~S.~Sathyaprakash}
\affiliation{\PSUPHYS}
\affiliation{\PSUASTRO}
\affiliation{\CARDIFF}
\author{Kentaro~Takami}
\affiliation{\KCCT}
\affiliation{\ITP}

\begin{abstract}
We show how gravitational-wave observations with advanced detectors of
tens to several tens of neutron-star binaries can measure the
neutron-star radius with an accuracy of several to a few percent, for
mass and spatial distributions that are realistic, and with none of the
sources located within $100\,{\rm Mpc}$. We achieve such an accuracy by
combining measurements of the total mass from the inspiral phase with those
of the compactness from the postmerger oscillation frequencies. For
estimating the measurement errors of these frequencies we utilize
analytical fits to postmerger numerical-relativity waveforms in the time
domain, obtained here for the first time, for four nuclear-physics
equations of state and a couple of values for the mass. We further
exploit quasi-universal relations to derive errors in compactness from
those frequencies. Measuring the average radius to well within 10\% is 
possible for a sample of 100 binaries distributed uniformly in volume 
between 100 and 300 Mpc, so long as the equation of state is not too soft or 
the binaries are not too heavy. 
\end{abstract}


\preprint{[LIGO-P1700071]}

\maketitle


\noindent\emph{Introduction.} The direct observation of gravitational
waves (GWs) by LIGO~\cite{Abbot2016-GW-detection-prl} has increased the
expectation that advanced GW detectors will also detect other types of
binaries, including binary neutron stars (BNSs). Imprinted in the emitted
GWs from such systems is the signature of the equations of state (EOSs)
of nuclear matter. For BNSs, this signature manifests itself during the
inspiral phase, when the two stars are tidally deformed, and in the
postmerger phase, when an unstable hypermassive neutron star (HMNS) can
form, emitting GWs at characteristic frequencies~\cite{Baiotti2016}. In
both cases, however, these imprints will be extremely small and the
accuracy of measurement of EOS parameter(s) will be poor, even in
detectors like Advanced LIGO (aLIGO)~\cite{Aasi:2014} and Advanced Virgo
(AdV)~\cite{Acernese:2014}, unless the binary happens to be nearby.

One way to address this problem is to combine the information in multiple
observations~\cite{Abadie:2010} with the expectation that the EOS
parameter errors will
reduce as the number of observations increases. For instance, the tidal
deformability parameter would typically go down as the
inverse-square-root of the number of BNS detections~\cite{DelPozzo2013,
  Agathos2015}. Yet, several tens of observations are needed to reduce
the errors to a level where only extreme EOSs can be distinguished. An
alternative method is to measure the characteristic frequencies of the
merger and postmerger signals~\cite{Bauswein2011, Stergioulas2011b,
  Takami2014, Takami2015, Bernuzzi2015a}; \eg the frequency at amplitude
maximum, $f_{\rm max}$, correlates closely with the tidal deformability
of the two stars \cite{Read2013, Bernuzzi2014, Takami2015}, and the
spectrum of the postmerger GW signal exhibits at least three strong peaks
of increasing frequency, dubbed $f_1,\,f_{2}$, and
$f_3$~\cite{Takami2014, Takami2015}.

In this \textit{Letter}, we explore how well the radius of a neutron star
can be measured by utilizing both the inspiral and postmerger phases of
the signal from multiple observations.  For this purpose, we utilize
numerical-relativity simulations to devise an analytical model of the
postmerger waveforms of four reference nuclear-physics EOSs (ALF2, SLy,
H4, and GNH3; see \cite{Takami2015} for details) in terms of a linear
superposition of damped signals with characteristic frequencies $f_1$ and
$f_2$. The model allows us to estimate errors $\Delta f_{1,2}$, which are
very large for individual observations in aLIGO or AdV as the
signal-to-noise ratio (SNR) of postmerger oscillations is $\lesssim\,1$
for a source at $\sim\,200\,{\rm Mpc}$. However, the joint error, e.g.,
in $f_2$, for a population of $\simeq\,100$ BNSs, uniformly distributed
in the comoving volume between $100\,{\rm Mpc}$ and $300\,{\rm Mpc}$, and
observed in the aLIGO-AdV three-detector network, is a few to several
percent, depending on the EOS. In essence, for a given binary with
average mass $\bar{M}$ and average radius at infinite separation
$\bar{R}$, the quasi-universal relations between characteristic
frequencies $f_{1}$ and $f_{2}$ and compactness ${\mathcal
  C}:=\bar{M}/\bar{R}$~\cite{Takami2015, Rezzolla2016} can be used to
deduce the error in $\mathcal{C}$ from the errors in those frequencies,
for various masses and mass ratios.\footnote{While our analysis utilizes
  these relations, it is not affected by how strictly universal they
  are.} Such measurement of $\mathcal{C}$ can be combined with that of
the total-mass from the inspiral to estimate the average radius for a BNS
population. We show that for these $\simeq\,100$ BNS observations the
error in radius is $2-5\%$ for stiff EOSs and $7-12\%$ for soft EOSs.
Our conclusion is that advanced detectors can help discriminate 
between stiff and soft EOSs. However, distinguishing two stiff EOSs, 
will be harder, with additional difficulties for very soft EOSs, whose 
postmerger signal is considerably weaker.

With important differences, our conclusions broadly agree with those
presented recently by other groups. Agathos~et~al.~\cite{Agathos2015}
estimated the evolution of the medians and $95\%$ confidence intervals in
the measurement of the leading-order term $c_0$ in the expansion of the
tidal deformability at the reference mass of $1.35\,M_\odot$, for some
reference EOSs in simulated aLIGO data~\cite{Agathos2015} and a
Gaussian mass distribution.
They found that
inspiral signals from $\approx 100$ or more BNSs are required for
determining $c_0$ to $10\%$ accuracy. Our analysis is different in that
instead of constructing Bayesian posteriors of $c_0$ from the inspiral
waveform, we use Monte-Carlo simulations to estimate the mean population
radius, but require similar number of sources for discriminating similar
pairs of EOSs.

Clark~et~al.~\cite{Clark2016} have instead used principal-component
analysis to infer the postmerger waveform in various planned or proposed
detectors and deduced that in aLIGO the radius of a BNS at a distance of
$30\,{\rm Mpc}$ and with component masses of $1.35\,M_\odot$ each can be
estimated to within $430\,{\rm m}$, which is a $3-4\%$ error. This result
appears to agree with our strong-signal case discussed below up to a
factor of two. However, their estimates of the postmerger amplitudes are
likely affected by the use of more dissipative numerical methods than
those employed here and by an approximate treatment of general
relativity. We also account for the deterioration in the measurement
arising from covariances of BNS masses and the postmerger frequencies
values, on the one hand, and the improvement in estimation accuracy that
can be had from knowledge of the total-mass from the inspiral
phase, on the other hand.

\noindent\emph{Postmerger waveforms.} Numerical-relativity simulations
have shown that the most likely product of a BNS merger is a metastable
HMNS that exists for several tens of milliseconds before collapsing to a
rotating black hole~\cite{Baiotti2016}. The GWs emitted from such an
oscillating, bar-shaped object show a strong correlation with the
stiffness of the nuclear material and hence with the
EOS~\cite{Baiotti2016}. Although also dependent on the total-mass, 
mass-ratio and EOS, the postmerger GW signal has robust spectral features with
prominent peaks at increasing frequencies $f_1,\,f_2,\,f_3$. These peaks
are reminiscent of spectral lines in atomic transitions, so that
imprinted in the spectrum of the postmerger signal is the state of dense,
nuclear matter. The analogy with atomic spectral lines is broader as it
is possible to infer cosmological redshift of a BNS merger from GW
observations alone, by measuring the Doppler shift in postmerger spectral
peaks of BNS mergers~\cite{Messenger2013}.

\begin{figure*}
\begin{center}
\includegraphics[width=1.9\columnwidth]{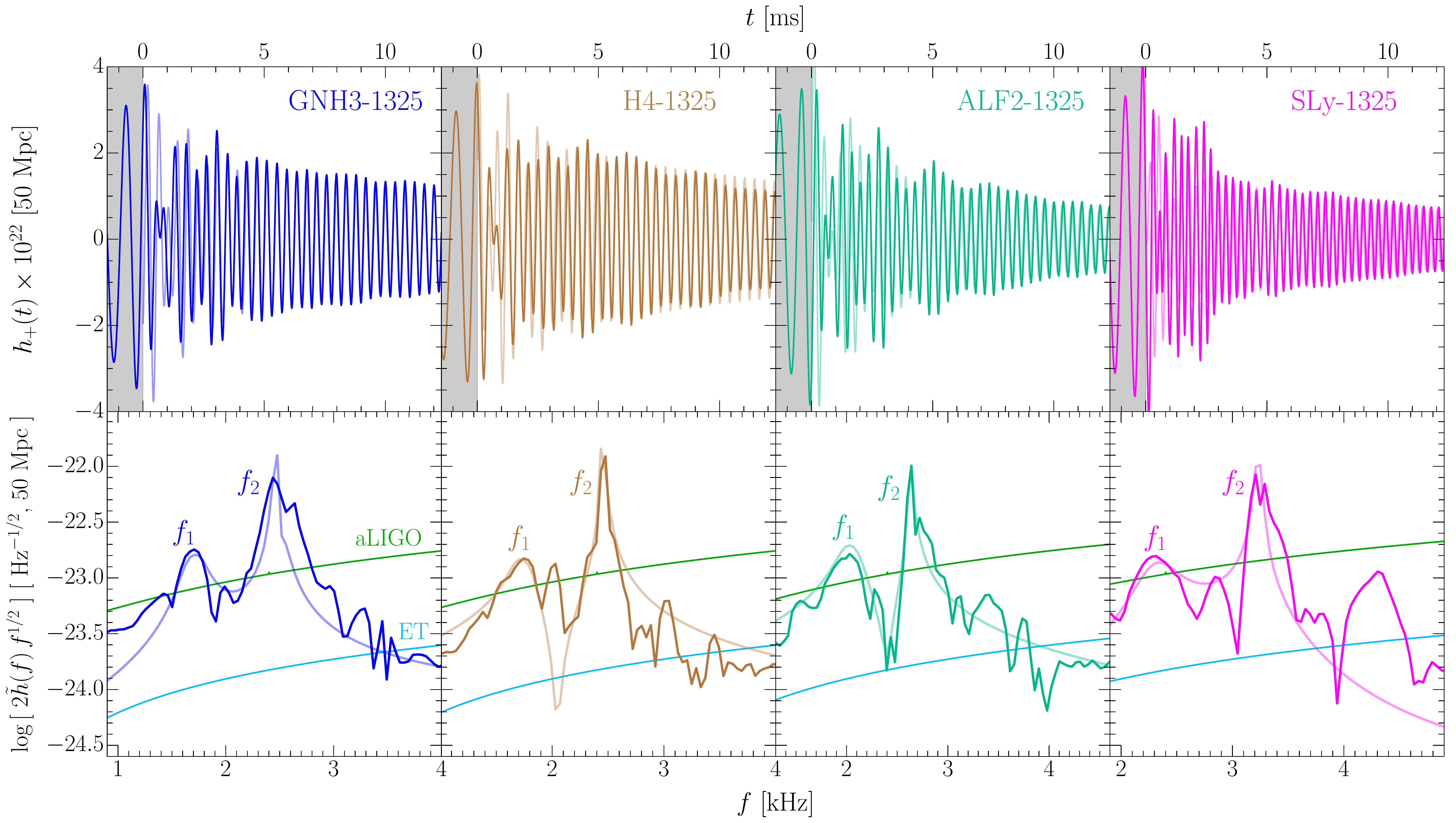}
\caption{\textit{Top panels:} Postmerger strain from numerical-relativity
  waveforms for four EOSs and a representative mass of
  $\bar{M}=1.325\,M_{\odot}$; our analytical ansatz is shown as a
  transparent line of the same color. Only the initial $12\,{\rm ms}$ of
  the complete $25\,{\rm ms}$ waveforms are reported to aid the
  comparison. \textit{Bottom panels:} Corresponding spectral amplitudes
  shown with the same color convention, superposed on the strain
  sensitivity curves of aLIGO and ET. Similarly good matches are produced
  also for $\bar{M}=1.250\,M_{\odot}$ (\cf Table~I and Fig.~3 in the
  supplemental material).}
\label{fig:waveform_2x4_M1325}
\end{center}
\end{figure*}

It is generally accepted that the most prominent peak, $f_2$ (see
Fig.~\ref{fig:waveform_2x4_M1325}), reflects the spin frequency of the
$m=2$-deformed HMNS, while the origin of the broader $f_1$ peak is still
under debate. The fact that the $f_1$ peak is short-lived, disappearing after
a few milliseconds, and is accompanied by a symmetric peak at even larger
frequencies $f_3\,\sim\,2\,f_2\,-\,f_1$, supports the interpretation that
it is a transient signal produced right after the merger by the damped
collisions of the two stellar cores (see~\cite{Takami2015,Rezzolla2016}
for a toy model).

Accurate modeling of waveforms from BNSs requires computationally
formidable numerical-relativity calculations. Since we are interested in
constraining EOS parameters with extensive Monte-Carlo simulations of
signals from $\simeq 100$ binaries with various EOSs in $\simeq 100$
noise realizations and average measurements over hundreds of BNS
population realizations,
it is clear that the accuracy and costs of the numerical-relativity
calculations need to be traded with a less accurate but computationally
efficient description of the waveforms. Hence, we derived a
phenomenological model for the postmerger waveform using analytical fits
in the time domain to a catalogue of numerical-relativity
waveforms~\cite{Takami2015,Rezzolla2016} that can be expressed as a
superposition of damped sinusoids with a time-evolving instantaneous
frequency~\cite{Clark2014,Clark2016}:
$h_{+}(t)=\alpha\exp(-t/\tau_{1})\bigl[\sin(2\pi\,f_{1}t)+\sin(2\pi(f_{1}-f_{1\epsilon})t)+\sin(2\pi(f_{1}+f_{1\epsilon})t)\bigr]+\exp(-t/\tau_{2})\sin(2\pi\,f_{2}t+2\pi\gamma_2t^2+2\pi\xi_2t^3+\pi\beta_2)$.
Here, $t=0$ refers to the merger time, $f_{1\epsilon}=50\,{\rm Hz}$, and
the ansatz reproduces all of the postmerger ``$+$'' polarization signals,
up to an overall amplitude; this is to be contrasted with the ansatz
considered in~\cite{Clark2014}, which models the waveforms only after the
amplitudes have decayed to half of the initial values\footnote{Better
  matches can be obtained by including more terms and parameters in the
  ansatz; however, the main effect of a less than perfect match is a
  lower SNR; see also Ref.~\cite{Hotokezaka2013c} for an alternative
  ansatz.}. The above fit not only agrees very well with the signal
spectra near $f_1$ and $f_2$, but also with the signal phase in the
time-domain, giving matches of $\sim 80 - 94\%$. Therefore, when combined
with a semi-analytical model of the inspiral waveform, \eg via a
post-Newtonian expansion with tidal corrections, the fitting ansatz gives
a complete analytic description of the signal from merging BNSs. The
above fit, parameterized by eight parameters (see Table~I in the
supplemental material), produces an accurate representation of the
waveform phase and a reasonably good description of its amplitude. The
top panels in Fig.~\ref{fig:waveform_2x4_M1325} show numerical-relativity
amplitude $h_{+}(t)$ and the analytical fits for four different EOSs and
for sources at $50\,{\rm Mpc}$. The bottom panels show the corresponding
spectral amplitudes, $2\sqrt{f}|\tilde h(f)|$, and the sensitivity curves of aLIGO and
the Einstein Telescope~\cite{Punturo2010b}. Here $\tilde h(f)$ is the
Fourier transform of $h_{+}(t)$, 

Two remarks are in order: First, the four EOSs chosen provide a good
coverage of the plausible range in stiffness of nuclear matter, but do
not represent very soft EOSs, such as APR4~\cite{Akmal1998a}. The
corresponding postmerger signal is much more complex
~\cite{Takami2015,Rezzolla2016}, with beats between different frequencies
not reproduced with our simple fitting ansatz. Second, our fits best
represent equal-mass systems and although the masses in observed binaries
do not differ significantly, it is unlikely that LIGO sources have mass
ratio $q=1$. Nevertheless, the quasi-universal relations used here 
continue to be valid also for systems with mass ratio
$q\,\gtrsim\,0.8$~\cite{Takami2015, Rezzolla2016}.

Our analytic waveforms also facilitate the interpretation of the
Monte-Carlo results described below in terms of the Fisher information
matrix parameter estimates, which broadly agree with the former (see
Table~I in the supplemental material), except for the
soft EOSs.  For a source even at $50\,{\rm Mpc}$, the postmerger signal
alone will be difficult to {\em detect} in an aLIGO detector. As an
example, the postmerger waveform of the H4 binary with average mass
$1.325\,M_{\odot}$ (H4-1325) has
$|2\tilde{h}(f)\,f^{1/2}|\,\simeq\,10^{-22}/\sqrt{\rm Hz}$ at
$f=f_2\,\simeq\,2470\,{\rm Hz}$, with the frequency bin-width being
$\delta f \sim 100\,{\rm Hz}$. The aLIGO noise amplitude at this
frequency is $S_h(f_2) \simeq 1.26\times 10^{-46}~{\rm Hz}^{-1}$, thus
yielding an ${\rm SNR}\,\simeq\,|2\tilde{h}(f)f^{1/2}|[\delta
  f/(f\,S_h(f) )]^{1/2}\,\simeq\,1.8$.

A small postmerger SNR, however, does not necessarily imply that the 
observations contain no information. Rather, small-SNR postmergers can provide
constraints if combined constructively over a population of such
signals. As an example, a Fisher-matrix analysis gives the $1-\sigma$ 
error in measuring $f_1$ and $f_2$ for a population of $100$ H4-1325 BNSs
at $100\,{\rm Mpc}$ with optimal sky-position and orientation to be
$\Delta f_1/f_1\,\simeq\,10\%$ and $\Delta f_2/f_2\,\simeq\,1\%$, or
$\Delta f_1\,\simeq\,177\,{\rm Hz}$ and $\Delta f_2\,\simeq\,27\,{\rm
  Hz}$ in a single aLIGO detector (see Table~I in the supplemental
material). Exploiting the quasi-universal relations between
$f_1,\,f_2$ and the compactness (see
the left two panels in Fig.~2 in the supplemental material), we can infer
the error in $\mathcal{C}$ through error propagation.  For the
aforementioned $100$ BNS observations, we deduce from the error in $f_2$
(which is much better measured than $f_1$) that the fractional error in
the measurement of the compactness is as small as
$\approx\,1.0\,\%$. Similar results are obtained for the other EOSs, and
masses and are listed in Table~I in the supplemental material. We have
also verified that other fitting expressions for $f_1,\,f_2$, \eg in
terms of fractional powers of $\mathcal{C}$, yield very similar errors in
the radius estimates derived below.

\noindent\emph{Radius measurement from a single BNS.} For the H4-1325 BNS
at $200\,{\rm Mpc}$, with an aLIGO-AdV network ${\rm SNR}=14$ for the
complete inspiral-merger-postmerger signal (after averaging over sky
locations and orientations, which reduces the SNR by a factor of $2.26$
relative to that for optimal sky location and
orientation~\cite{Abadie:2010,Ghosh:2013yda}), the $1-\sigma$ measurement errors are
$\Delta f_2/f_2 \approx 14.0\%$ (Table~I in the supplemental material),
and derived from it via quasi-universal relations, $\Delta \mathcal{C}/\mathcal{C}\approx\,9\%$. Taking
the component mass $1-\sigma$ error to be
$11\%$~\cite{Rodriguez:2013oaa}, the error in radius from error
propagation is $\approx 14\%$.  For the same source at $30\,{\rm Mpc}$
with optimal location and orientation, the complete-waveform network-SNR
will be $\approx\,211$, even though the postmerger signal will have ${\rm
  SNR}\approx\,6.4$. At such a distance, the error in average binary mass
is much smaller, at $0.08\%$, and
$\Delta\,\mathcal{C}/\mathcal{C}\,\approx\,0.9\%$. In this strong-signal
case, the radius error reduces to $0.9\%$, or $125\,{\rm m}$. In a single
aLIGO detector, the error will rise to $\approx 215\,{\rm m}$. This is
roughly two times more accurate than the value given in
Ref.~\cite{Clark2016}, the primary reason being that their waveforms are
more rapidly damped than ours, as noted above.  Furthermore, while our
errors are estimated for the average radius of the parent BNS, the error
in Ref.~\cite{Clark2016} is estimated for the radius of a cold
nonrotating neutron star of mass $1.6\,M_\odot$ ($R_{1.6}$) and for a
single value of the average mass ($\bar{M}=1.350\,M_{\odot}$); we find
this approach not applicable to our data and that of other groups (see
Fig.~5 in the supplemental material). Finally, other constraints can be
imposed from the quasi-universal relation that exists between the stellar
compactness and the tidal deformability parameter $\kappa_2^{T}$ (see
Fig.~1 in the supplemental material), which introduces additional
quasi-universal relations between $f_1,\,f_2$ and $\kappa_2^{T}$.

\noindent\emph{Radius measurement from a BNS population.} At such small
SNRs it is not possible to measure $f_{1,2}$ accurately. However, for a
population of $N>1$ BNSs it is possible to align and stack the $f_2$
peaks, so that for a large enough $N$, and uncorrelated noise across
those $N$ observations, the stacked amplitude spectra can have enough SNR
to allow for an accurate measurement of $f_2$. A realistic population
will have a variety of mass pairs, but since the total-mass of a BNS
system correlates well with $f_2$~\cite{Bauswein2012, Rezzolla2016}, one
can use a measurement of $M_{\rm tot} = 2 \bar{M}$ from the inspiral
waveform to deduce it.  To test this idea, we performed a Monte-Carlo
simulation (see supplementary materials) comprising multiple time-series,
each with a simulated postmerger signal from this BNS population added to
Gaussian noise with aLIGO zero-detuned-high-power noise power-spectral
density~\cite{aLIGOZDHP:web}.  
Similar to Ref.~\cite{Clark2016}, we rescaled the multiple
signal spectra to align the $f_2$ values deduced from the (generally erroneous)
total-mass estimate for each signal to stack all at a chosen common
frequency, $f_2^c$.  Standard spectral frequency estimation yielded the
value of $f_2^c$ and its statistical spread for that population.  We next
used the quasi-universal relation between $f_2$ and compactness, and
error-propagation, to deduce the error in the average neutron-star radius
of that population.
 
In our Monte-Carlo simulations, each experiment employed $100$ BNS
postmerger signals injected in 100 uncorrelated time-series of Gaussian
noise with aLIGO zero-detuned-high-power noise power-spectral
density. The stacking method described above was used to deduce $f_2^c$
for this experiment. We repeated this hundred times by changing only the
noise realizations and found the mean $f_2^c$, and its $90\%$ confidence
interval, for this experiment ensemble. Finally, to keep error
fluctuations to less than a percent, we computed the average
confidence interval over 900 copies of experiment ensembles by changing
the BNS sources (within the chosen range of masses and distances) and
noise realizations.

Since the mass distribution of extragalactic BNSs is not known, we study
two different populations. In the first case we took the masses to be
uniformly distributed in a range listed below.  In the second case, we
built a large set of normally distributed masses centered at
$1.35\,M_\odot$, with standard-deviation $0.05\,M_\odot$, to mimic the
masses in galactic BNSs~\cite{Agathos2015}. We then drew our sample
of $2N$ masses from this distribution by restricting them to lie within a
given range.

\begin{figure}
\begin{center}
\includegraphics[width=0.45\textwidth]{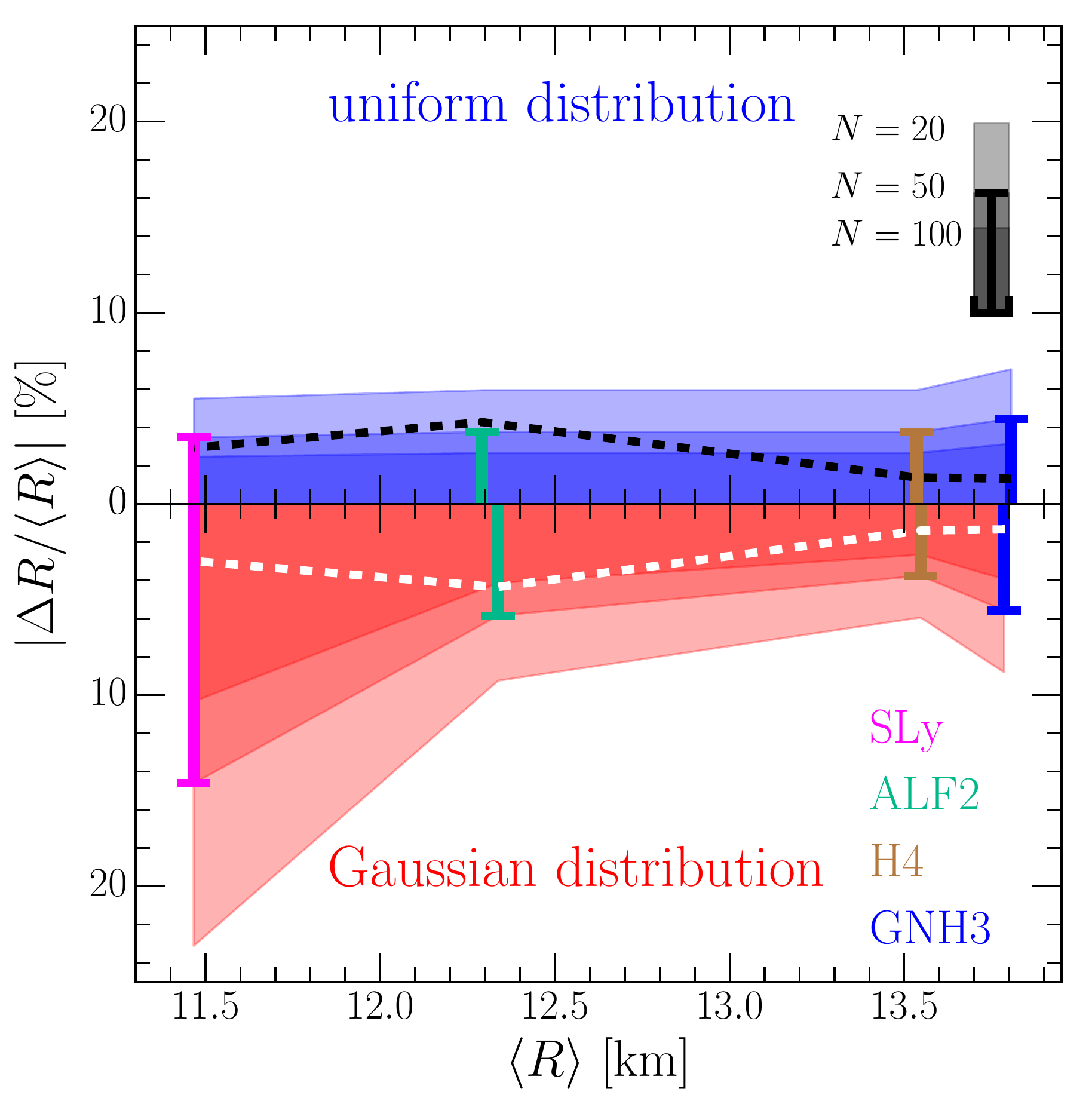}
\caption{Estimated relative error in the radius measured, at 90\%
  confidence level, versus the average population radius for different EOSs and
  $N=20,\,50,\,100$ (different shadings) BNSs distributed uniformly in a
  comoving volume between $100$ and $300\,{\rm Mpc}$. The two panels refer
  to binaries whose distribution in mass in the range
  $[1.2,\,1.38]\,M_\odot$ is either uniform (top) or Gaussian
  (bottom). Shown with dashed lines are the errors from the Fisher-matrix
  analysis for $N=50$.}
\label{fig:deltaR_vs_R}
\end{center}
\end{figure}

For all EOSs and the two mass distributions (Gaussian and uniform) the
radius errors found from Monte-Carlo studies are similar to those
obtained from Fisher studies, provided one limits the masses to the range
$[1.2,\,1.38]\,M_\odot$ (see Fig.~\ref{fig:deltaR_vs_R}); the notable
exception is the error for the Gaussian distribution with the SLy
EOS. The reason for the agreement with the Fisher-matrix estimates is
that the average value of $f_2$ is not very high. However, for the
Gaussian mass distribution for SLy, the average $f_2$ is the highest, so
that for the same percentage error in $f_2$, the error $\Delta f_2$ is
largest for SLy. This implies that the stacking of signals works less
perfectly and the summed signal at the fiducial frequency grows slower
with the number of observations than what is realized in the Fisher
method. To confirm this behaviour, we considered two sets of Monte-Carlo
simulations, one with the same population of $100$ BNSs but with all
masses set to $1.25\,M_\odot$; in the second one we took the masses to be
$1.325\,M_\odot$. For the SLy EOS, we found that the radius error is
about $2.7\%$ for the first (low-mass) case, but rises to about $10\%$
for the second (high-mass) case,  at 90\% confidence level.

From Fig.~\ref{fig:deltaR_vs_R} it is clear that as the EOS gets softer,
the Fisher-matrix errors will be less credible and that if the EOS turns
out to be soft, then measuring the radius to an accuracy of $10\%$ will
be challenging with aLIGO-like detectors.

\noindent\emph{Conclusions.} We have presented a new method to infer the
average radius of a population of neutron stars in BNSs that employs
both the inspiral-merger and the postmerger phases, and that may provide
us with a useful alternative to the one where it is inferred from the
tidal corrections to the inspiral-merger phase alone. In particular, the
postmerger phase allows for the measurement of the compactness, which
nicely complements the measurement of the component masses via the
inspiral phase to help us determine the radius. Hence, our
phenomenological time-domain modeling of the postmerger waveform can
prove handy for the production of complete inspiral, merger and
postmerger time-domain waveforms.

It may be argued that our results are somewhat limited for at least two
reasons. First, our phenomenological fits and the estimates of the errors
$\Delta\,f_{1,2}$ are given for binaries with mass ratio
$q\simeq\,1$. However, we have found that similar fits can be obtained
for unequal mass-ratios studied in~\cite{Rezzolla2016}, and that
$\Delta\,f_{1,2}$ are very similar in such cases for signals with the
same SNRs. This observation is consistent with those made in
\cite{Clark2016}. If nature relents to provide us with an especially
strong signal, such that the network SNR of the postmerger signal is
$\approx 6.4$, which can happen if the source is of optimal orientation and
sky-position, and located at a distance of $30\,{\rm Mpc}$, then our
method can be used to deduce the radius to about $1.6\%$,  
at 90\% confidence level. Second, as the
number of observed binaries increases and the fractional errors of the
EOS properties decrease, the systematic uncertainties, mostly related to
the accuracy of numerical-relativity calculations, will become the
dominant source of errors. While this is somewhat inevitable as the
simulations are still too expensive to provide Richardson-extrapolated
results, or to include the most sophisticated treatments of
magnetic-field growth and neutrino emission, progress is continuously
made on this front and the results that will be obtained in the coming
decade will significantly reduce the impact of these systematic
uncertainties.

Finally, since both the imprint of EOS and the signals themselves may be
weak, it will be important to utilize as much of the signal as is
meaningful for measuring the EOS parameters. This can be especially
helpful owing to the possibility that these parameters may have
non-trivial covariances with other parameters, such as their masses. EOS
estimation would therefore gain from exploring if the same EOS parameter
values can explain consistently features in all parts of the waveform,
specifically, the inspiral and the postmerger waveforms.


\smallskip\noindent\emph{Acknowledgements.} It is a pleasure to thank
J. Clark, B. Lackey and J. Read for reading the manuscript and providing
useful input. Support comes from: NSF grants (PHY-1206108, PHY-1506497);
ERC Synergy Grant ``BlackHoleCam'' (Grant 610058); ``NewCompStar''; COST
Action MP1304; LOEWE-Program in HIC for FAIR; European Union's Horizon
2020 Research and Innovation Programme (Grant 671698) (call
FETHPC-1-2014, project ExaHyPE); JSPS KAKENHI grant (Grant 15H06813,
17K14305); and the Navajbai Ratan Tata Trust. The simulations were
performed on SuperMUC at LRZ-Munich, on LOEWE at CSC-Frankfurt and on
Hazelhen at HLRS in Stuttgart. This manuscript was assigned the LIGO
Document No. LIGO-P1700071.


\bibliographystyle{apsrev4-1}

\newpage
\appendix

\section*{Supplemental material}
The purpose of this Appendix is to summarize the existence of the 
quasi-universal relations that have been employed in the Fisher 
matrix analysis and the Monte-Carlo simulations 
described in the main text. Although most of the
information on these relations has been presented in a number of 
previous publications~\cite{Takami2014, Takami2015, Rezzolla2016}, 
it is useful to summarize it here.

We start by recalling some definitions. The tidal deformability parameter
$\kappa_2^{^T}$ for a generic unequal-mass binary is defined as
\begin{equation}
\label{eq:kappa2T}
\kappa_2^{T} :=
2\left[
         q \left(\frac{X_{_A}}{\mathcal{C}_{_A}}\right)^5k^{^A}_2 +
\frac{1}{q}\left(\frac{X_{_B}}{\mathcal{C}_{_B}}\right)^5k^{^B}_2\right]\,,
\end{equation}
where $A$ and $B$ refer to the primary and secondary stars in the binary,
\begin{align}
&q:=\frac{M_{_B}}{M_{_A}} \leq 1\,, &
&X_{_{A,B}}:=\frac{M_{_{A,B}}}{M_{_A}+M_{_B}}\,, &
\end{align}
$k_2^{^{A,B}}$ are the $\ell=2$ dimensionless tidal Love numbers, and
$\mathcal{C}_{_{A,B}}:=M_{_{A,B}}/R_{_{A,B}}$ are the compactnesses. In
the case of equal-mass binaries, $k_2^{A}=k_2^{B}=\bar{k}_2$, and
expression \eqref{eq:kappa2T} reduces to
\begin{equation}
\kappa_2^{^T}:=\frac{1}{8}\bar{k}_2
\left(\frac{\bar{R}}{\bar{M}}\right)^5=\frac{3}{16}\Lambda=
\frac{3}{16} \frac{\lambda}{\bar{M}^5} \,,
\end{equation}
where the quantity
\begin{equation}
\lambda:=\frac{2}{3} \bar{k}_2 {\bar{R}}^5,
\end{equation}
is another common way of expressing the tidal Love number for equal-mass
binaries~\cite{Read2013}, while $\Lambda:=\lambda/\bar{M}^5$ is its
dimensionless counterpart and was employed in~\cite{Takami2015}.

\begin{figure}
\begin{center}
\includegraphics[width=0.4\textwidth]{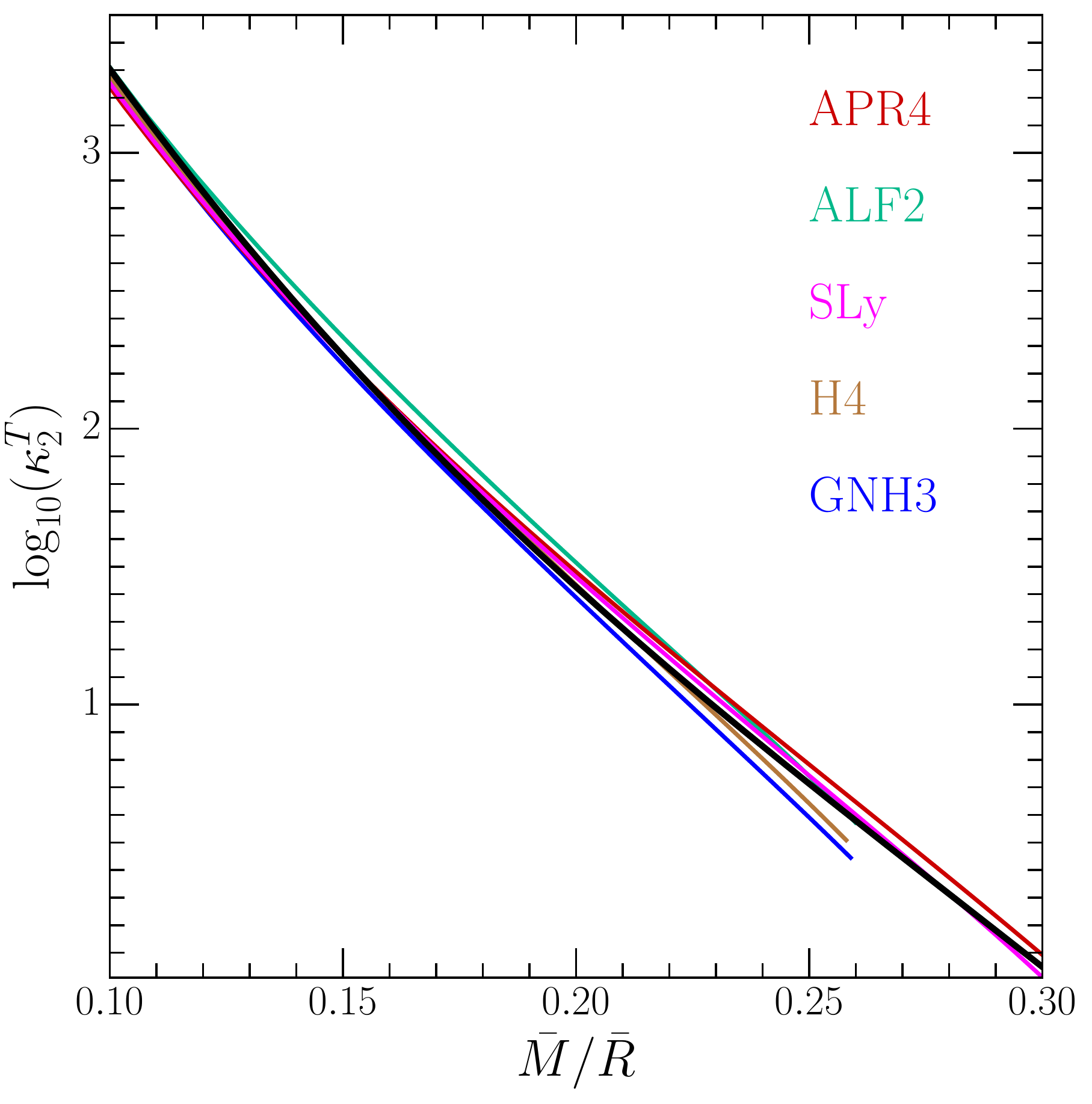}
\caption{Quasi-universal relation between the tidal deformability
  parameter and the stellar compactness for the four EOSs considered
  here and for APR4. Indicated with a black solid line is the fit given by
  Eq.~\eqref{eq:kappa_vs_C}.}
\label{fig:C_vs_kappa}
\end{center}
\end{figure}
A quasi-universal relation, which is also present at the level of
equilibrium solutions of nonrotating models, is the one relating the
tidal deformability and the stellar compactness shown in
Fig.~\ref{fig:C_vs_kappa} for the five EOSs considered here. It is not
difficult to express this rather tight correlation with a polynomial of
the type
\begin{equation}
\label{eq:kappa_vs_C}
\log_{10}\left(\kappa^{T}_2\right) \simeq d_0 + d_1\, \mathcal{C} + d_2\,
\mathcal{C}^2 + d_3\, \mathcal{C}^3\,,
\end{equation}
where $d_0=6.29,\,d_1=-37.41,\,d_2=85.68,\,d_3=-101.07$ for stable models
with $\mathcal{C}>0.05$. It is shown as a black solid line in
Fig.~\ref{fig:C_vs_kappa}. Note that this relation is valid over a range
of compactness that is much larger than the one considered in the
analysis here; more importantly, it effectively represents a way to map
any measured quantity expressed in terms of $\kappa^{T}_2$ to a quantity
expressed in terms of $\mathcal{C}$, and {\it vice versa}.
Specifically, it introduces additional quasi-universal relations
between frequencies~\cite{Takami2015, Rezzolla2016},
\begin{align}
&f_1\,\approx\,c_0+c_1\,x+c_2\,x^2+c_3\,x^3 && {\rm kHz}\\
&f_2\,\approx\,5.832-1.118\,x && {\rm kHz}\,,
\end{align}
where $x:=(\kappa_2^T)^{1/5},\,c_0=45.195,\,c_1=-43.484,\,c_2=14.653$ and
$c_3=-1.662$.

\begin{figure*}[bt]
\begin{center}
\includegraphics[width=0.235\textwidth]{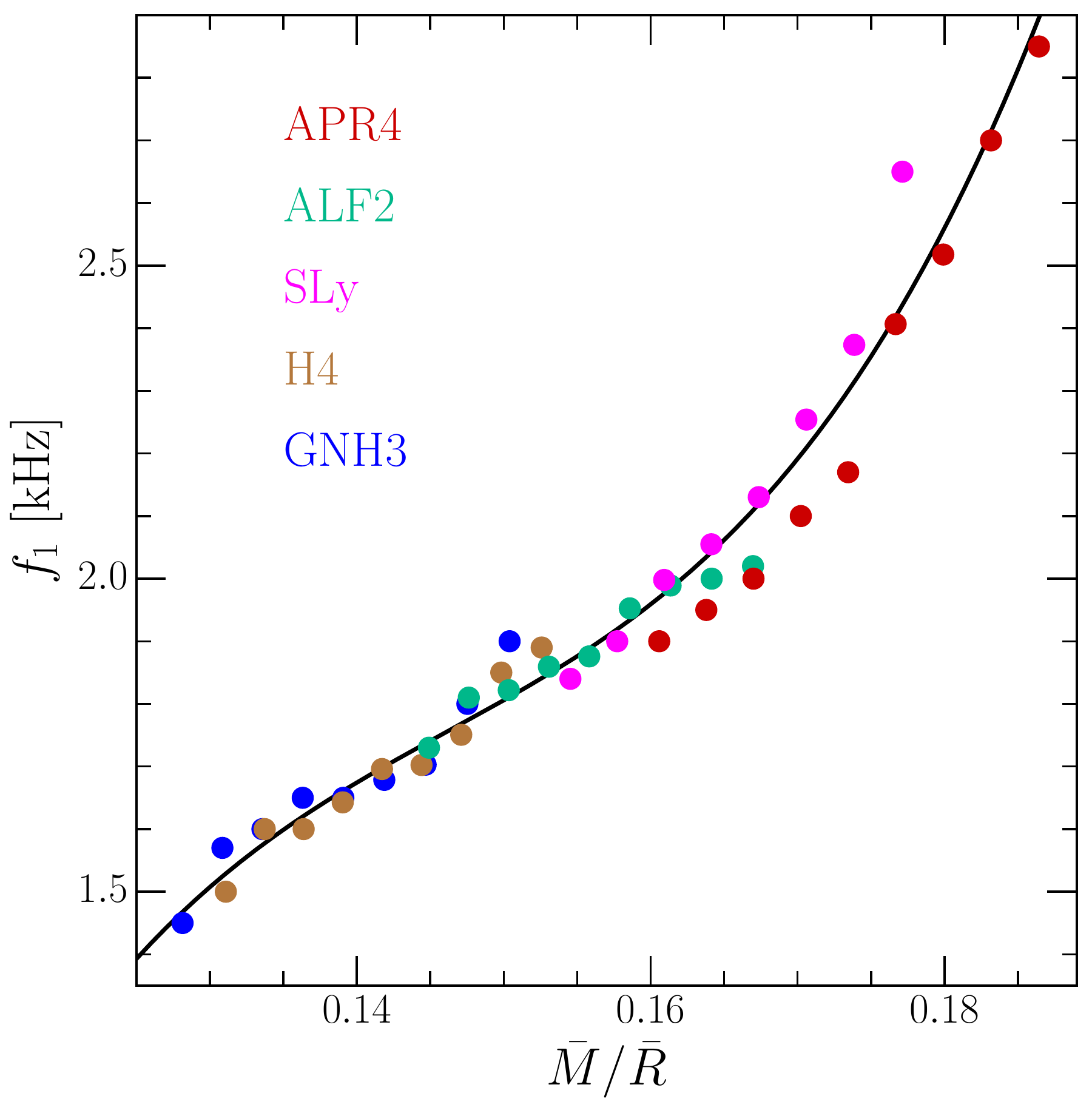}
\hskip 0.25cm
\includegraphics[width=0.235\textwidth]{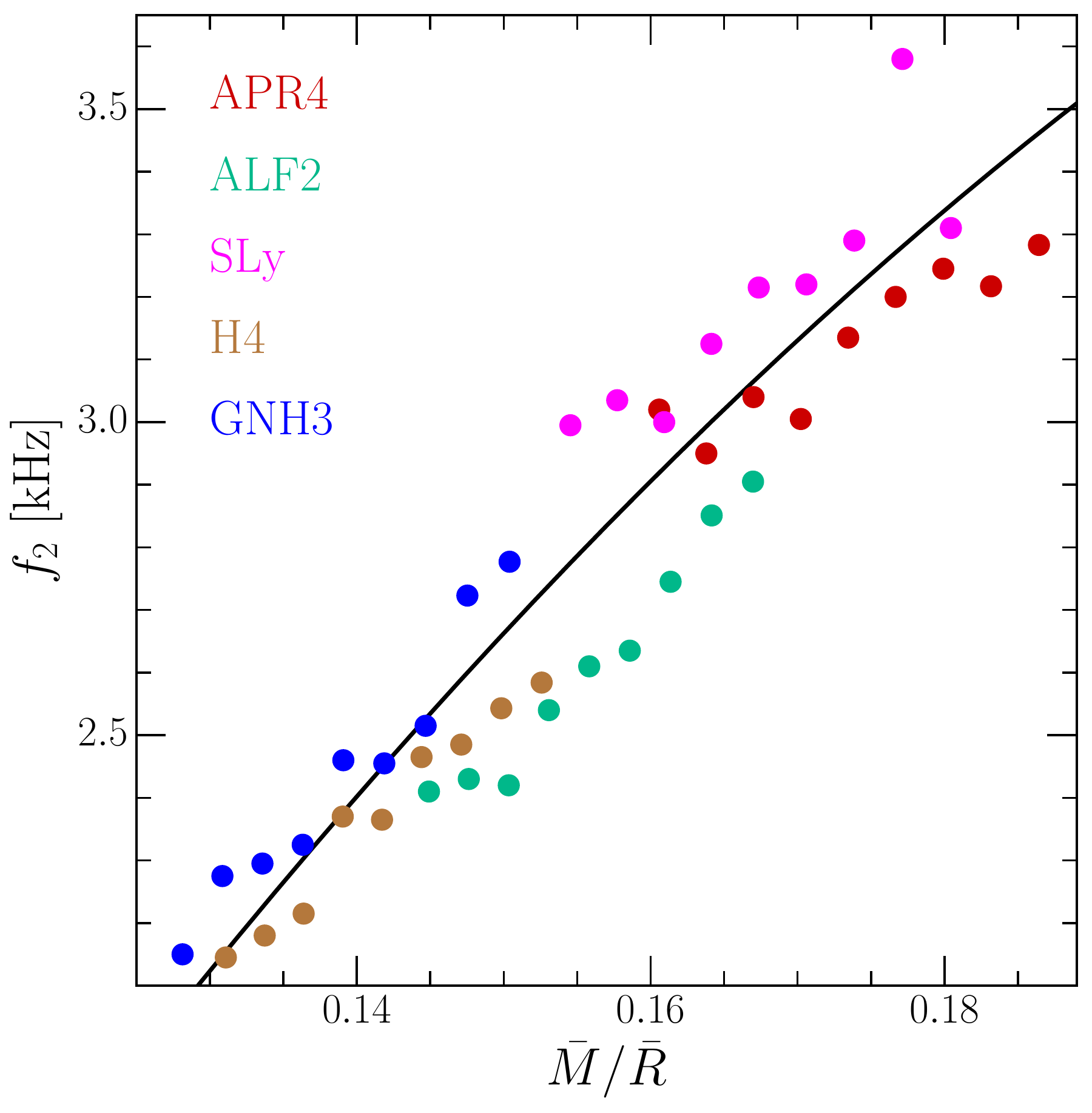}
\hskip 0.25cm
\includegraphics[width=0.235\textwidth]{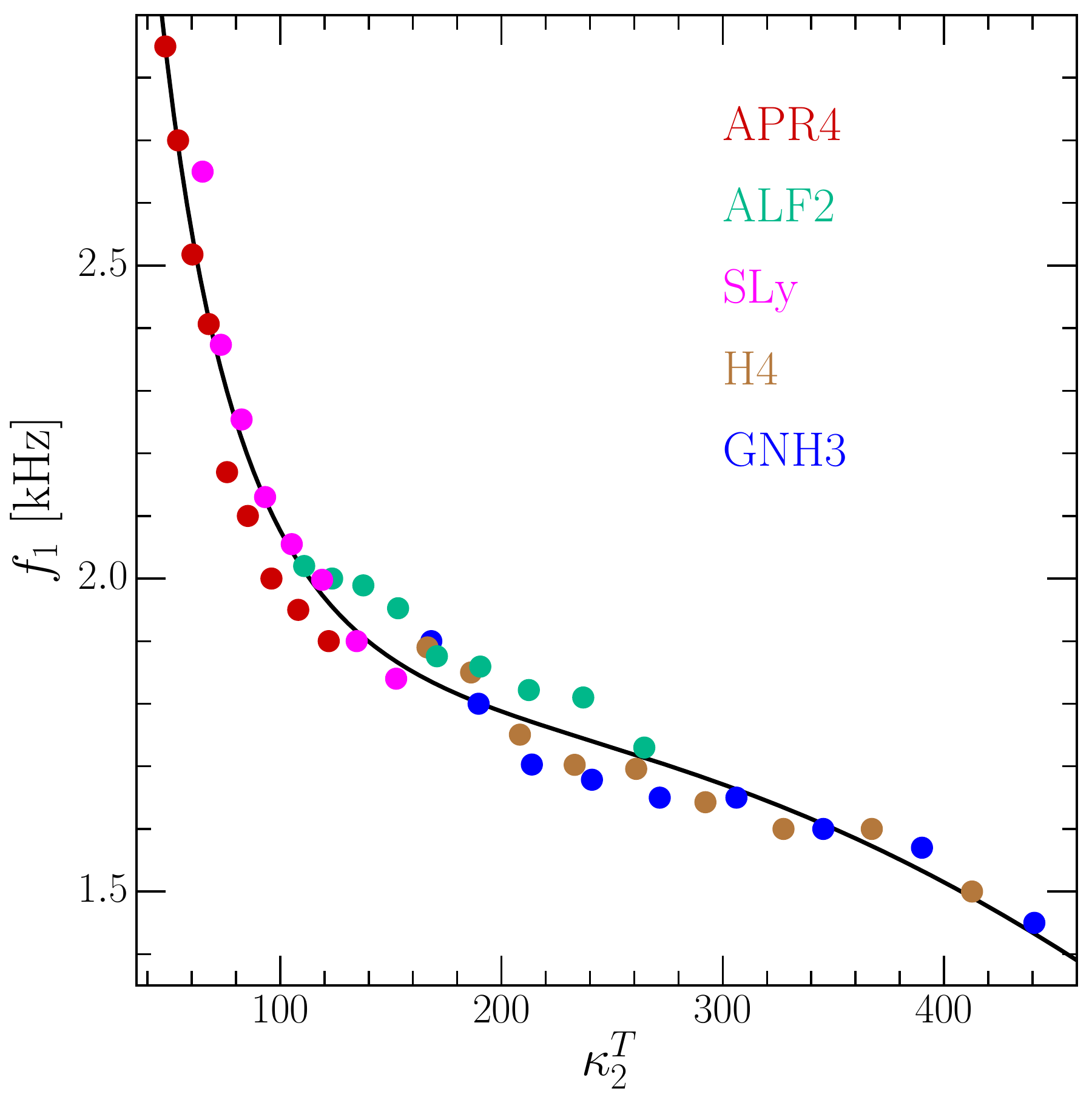}
\hskip 0.25cm
\includegraphics[width=0.235\textwidth]{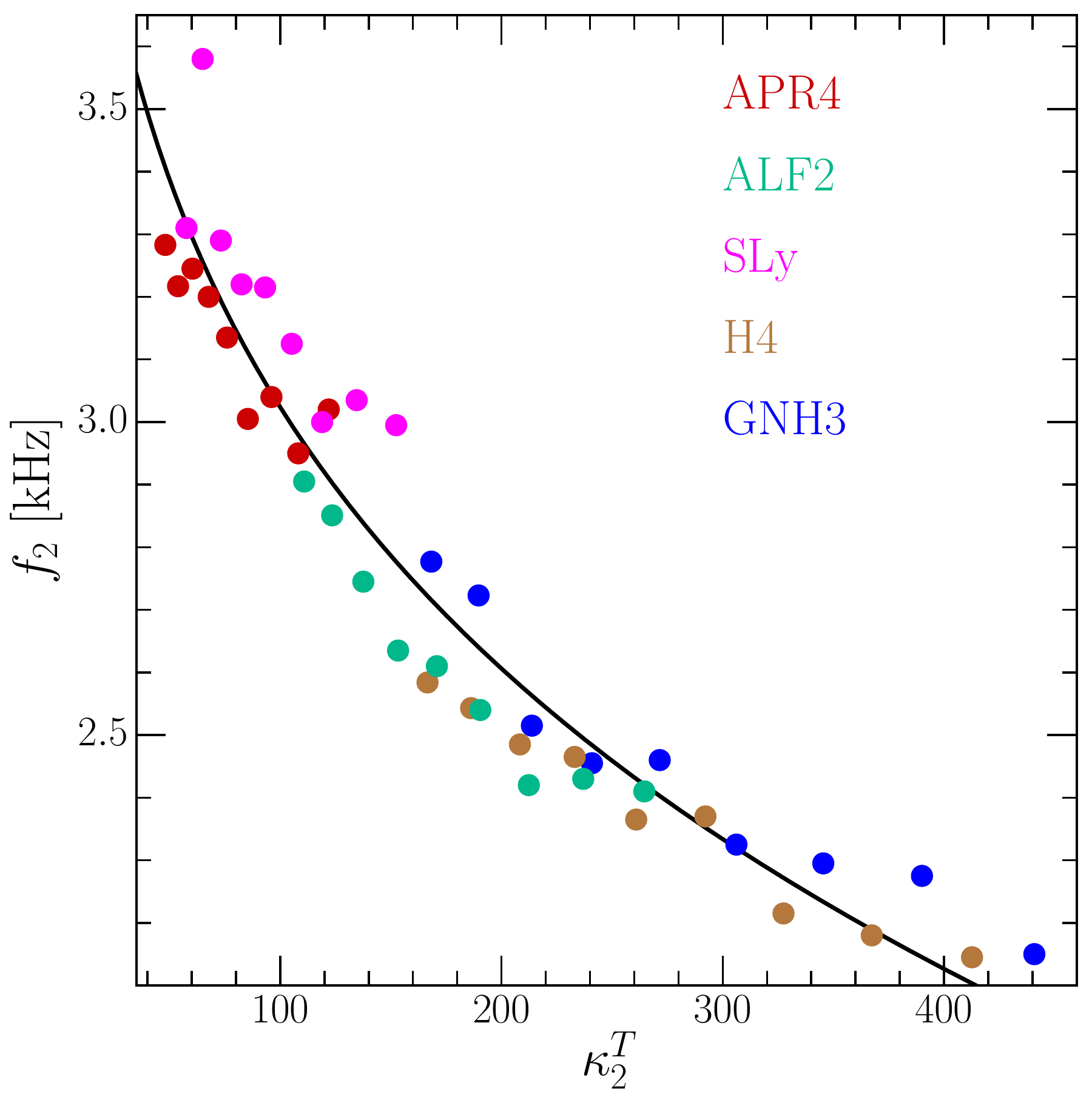}
\caption{\textit{Left two panels}: Quasi-universal behavior of the
  $f_1,\,f_2$ frequencies when expressed as a function of the average
  compactness $\mathcal{C}=\bar{M}/\bar{R}$ of the neutron stars
  comprising the binary. Filled circles of different colors refer to
  equal-mass binaries with different cold EOSs. The black solid line
  shows the analytic fit. \textit{Right two panels}: The same as in the
  left two panels but when $f_1,\,f_2$ are expressed as a function of the
  tidal deformability parameter $\kappa_2^T$.}
\label{fig:f12_vs_C}
\end{center}
\end{figure*}

Figure \ref{fig:f12_vs_C} shows instead the quasi-universal behavior
found numerically. The left two panels refer to the $f_1,\,f_2$
frequencies when expressed as a function of the average compactness
$\mathcal{C}=\bar{M}/\bar{R}$ of the neutron stars comprising the
binary. Filled circles of different colors refer to equal-mass binaries
with different EOSs. Black solid lines correspond to the analytic
fits~\cite{Takami2015,Rezzolla2016}:
\begin{align}
&f_1\,\approx\,a_0+a_1\,\mathcal{C}+a_2\,\mathcal{C}^2+a_3\,\mathcal{C}^3
&& {\rm  kHz}\\
&f_2\,\approx\,b_0+b_1\,\mathcal{C}+b_2\,\mathcal{C}^2 && {\rm kHz}\,,
\end{align}
where
$a_0=-35.17,\,a_1=727.99,\,a_2=-4858.54,\,a_3=10989.88,\,b_0=-3.12,\,b_1=51.90$
and $b_2=-89.07$.  The two right panels show $f_1,\,f_2$ as a function of
the tidal deformability parameter $\kappa_2^T$. As before, the black
solid line shows the analytic fit. For a GW signal with a complete
waveform ${\rm SNR} \simeq 14$, the inspiral phase will determine
$\bar{M}$ with a $1-\sigma$ accuracy of about
$11\,\%$~\cite{Rodriguez:2013oaa}, which is better than how accurately
the same parameter can be measured from the postmerger phase
alone. Therefore, the fitting functions for $f_1$ and $f_2$ show that the
error in measuring $f_1$ from observations of the postmerger phase can
improve the accuracy with which the EOS parameter $\lambda$ can be
determined.

\begin{table*}[h]
\begin{tabular}{lcccccccccccc}
\hline
binary & $f_1$    & $\tau_{1}$       & $f_2$
& $\tau_{2}$      & $\gamma_2$    & $\xi_2$       & $\alpha$ 
& $\Delta f_1$    & $\Delta f_2$    & $\Delta\mathcal{C}/\mathcal{C}$ & $\Delta f_2^{\rm MC}$   & [$\Delta R/R]^{\rm MC}$   \\
& $[\mathrm{kHz}]$ &$[\mathrm{ms}]$ &
$[\mathrm{kHz}]$ & $[\mathrm{ms}]$ & $[\mathrm{Hz}^2]$ &
$[\mathrm{Hz}^3]$  & & $[\mathrm{Hz}]$ &
$[\mathrm{Hz}]$ & [\%]  & $[\mathrm{Hz}]$ & [\%]    \\ \hline 
{GNH3-1250}  & 1.60 & 2  &  2.30  & 23.45 &    38 & -9.e2
& 0.46 &         371  &  29     & 1.0   &   14.3   &  1.8     \\ 
{H4-1250}    & 1.65 & 5  &  2.22  & 20.45 &  -677 &  0.0   & 0.55 &         151  &  43     & 1.2   &  50    &  2.7      \\ 
{ALF2-1250}  & 1.85 & 15 &  2.42  & 10.37 & -3467 &  2.e4  & 0.55 &          66  & 133 & 3.4    &   62.5   &  3.0     \\ 
{SLy-1250}   & 2.30 & 1  &  3.00  & 13.59 &     0 &  0.0   & 0.50 & 1683 &  82     & 2.2    &  52.0    &   2.4    \\ \hline 
{GNH3-1325}  & 1.70 & 2  &  2.45  & 23.45 &   342 &  5.e4
& 0.35 &         371  &  40     & 1.0    &  100    &  4.5      \\ 
{H4-1325}    & 1.75 & 5  &  2.47  & 20.45 & -1077 &  4.5e3 & 0.30 &         177  &  27     & 1.0    &  50    &  2.7     \\ 
{ALF2-1325}  & 2.05 & 15 &  2.64  & 10.37 &  -863 &  2.5e4 & 0.50 &          79  &  60     & 1.6    &   97   &  4.0     \\ 
{SLy-1325}   & 2.30 & 1  &  3.22  & 13.59 &  -617 &
5.5e4 & 0.50 & 1137 &  74     & 2.0    &   312   &  9.8      \\ \hline 
\end{tabular}
\caption{Parameter values for the analytic waveform models. $\beta_{2}$
  is adjusted to match numerical-relativity waveforms. $\Delta f_{1,2}$
  are Fisher-matrix $1-\sigma$ error estimates from postmerger signals in a single
  aLIGO detector for a reference population of $100$ binaries with
  optimal orientation and sky position, a distance of $200\,{\rm Mpc}$,
  and an integration time of $25\,{\rm
    ms}$. $\Delta\,\mathcal{C}/\mathcal{C}$ is deduced from $\Delta f_2$
  by error-propagation using quasi-universal relations. The errors in
  $f_2$ and the average radius, at 90\% confidence level, 
  of a BNS population with identical
  component masses and EOS, distributed uniformly in volume between
  100~Mpc and 300~Mpc, and averaged over orientation and sky position,
  obtained from Monte-Carlo simulations are listed in the last two
  columns, respectively.}
\label{fitsAllEOSs}
\end{table*}

\begin{figure*}
\begin{center}
\includegraphics[width=0.9\textwidth]{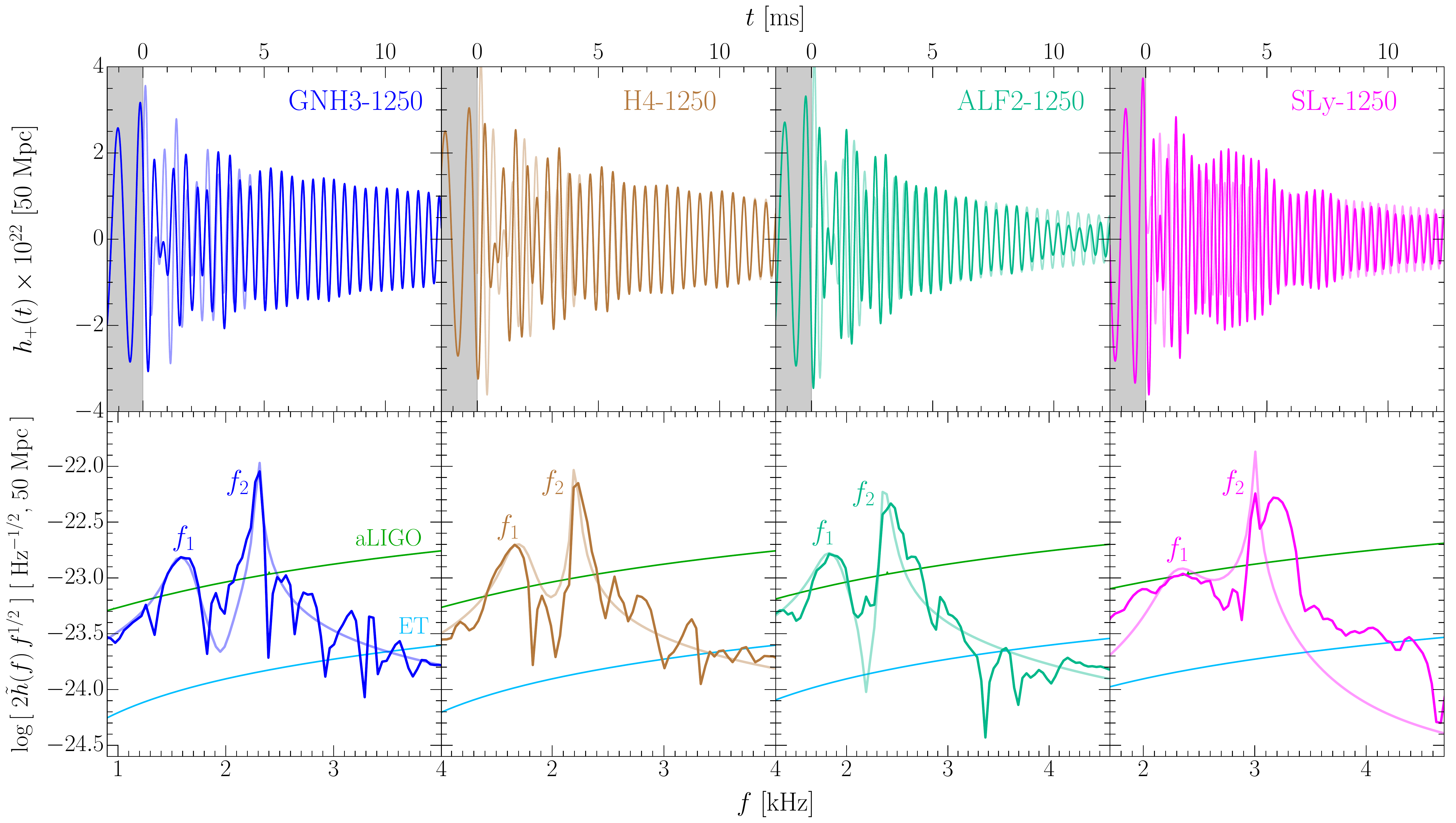}
\caption{The same as Fig.~1 in the main text, but for binaries with lower
  masses, \ie $M=2\times 1.250\,M_{\odot}$. In this case too the match
  between the numerical waveforms and the analytic ansatz is very good;
  even better matches are expected if the analytic ansatz is extended to
  model also the $f_3$ frequency peak.}
\label{fig:waveform_2x4_M1250}
\end{center}
\end{figure*}

Figure~\ref{fig:waveform_2x4_M1250}, which is analogous to Fig.~1 in the
main text, shows numerical waveforms and corresponding analytic fits for
a series of low-mass binaries, \ie $M=2\times 1.250\,M_{\odot}$. Note
that in this case too, and apart from the short transient right after the
merger (\ie for $0\lesssim t \lesssim 3\,{\rm ms}$) the match between the
numerical waveforms and the analytic ansatz is very good, both in the
time and frequency domains. Obviously, even better matches can be
obtained if a better modelling is used for the transient phase (\eg
following the mechanical toy model discussed in Appendix A
of~\cite{Takami2015}) or if the analytic ansatz is extended to include
also very high-frequency components, namely, by including the modelling
of the $f_3$ frequency peak.

\begin{figure}
\begin{center}
\includegraphics[width=0.4\textwidth]{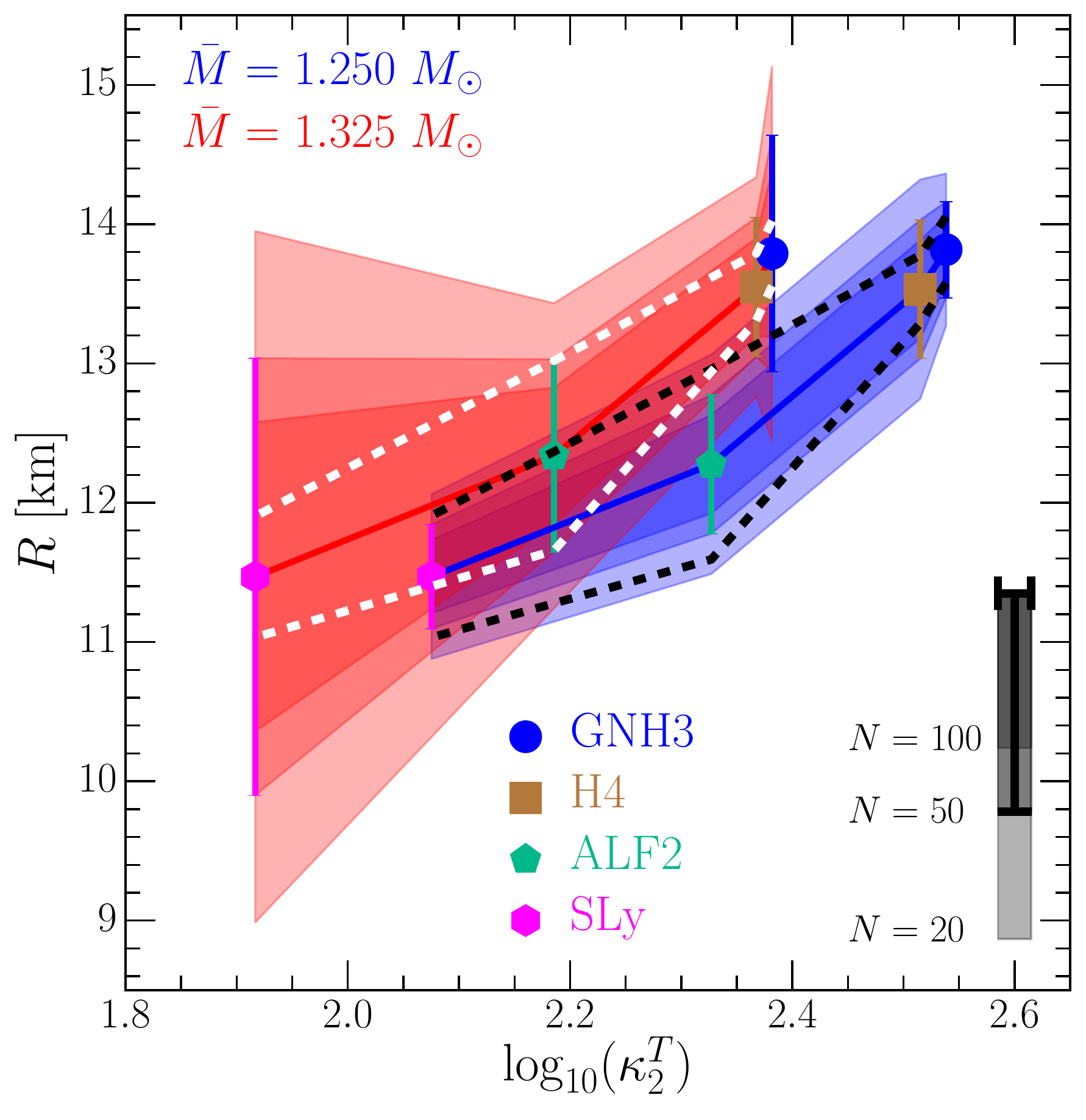}
\caption{Radius measurement error, at 90\% confidence level, 
  as a function of the tidal
  deformability when the population of BNSs, distributed uniformly in a
  comoving volume between $100$ and $300\,{\rm Mpc}$,
  is characterised by a
  single value of the average mass, \ie either $\bar{M}=1.250\,M_{\odot}$
  or $\bar{M}=1.325\,M_{\odot}$, for the four EOSs studied here.
  Different values of shading refer to the different numbers of binaries
  considered, \ie $N=20,\,50,\,100$. Note that soft EOSs have
  systematically larger uncertainties. Shown with dashed lines are the errors from the Fisher-matrix analysis for $N=50$.}
\label{fig:error_R}
\end{center}
\end{figure}

Finally, Fig.\,\ref{fig:error_R}, complementary to Fig.~2 in the main
text, shows errors in the reconstructed radii when the population of BNSs
is characterised by a single value of the average mass, \ie either
$\bar{M}=1.250\,M_{\odot}$ or $\bar{M}=1.325\,M_{\odot}$, for four
different EOSs. Different values of shading refer to different number of
binaries considered, \ie $N=20,\,50,\,100$.

The errors clearly grow as a function of the tidal deformability. In view
of the discussion relative to Fig.\,\ref{fig:C_vs_kappa} and the
quasi-universal relation \eqref{eq:kappa_vs_C}, Fig.\,\ref{fig:error_R}
essentially highlights a basic but important result: stellar radii
measurements will be more accurate for stiff EOSs, while they will
systematically suffer from larger uncertainties for soft EOSs.
\begin{figure}
\begin{center}
\includegraphics[width=0.4\textwidth]{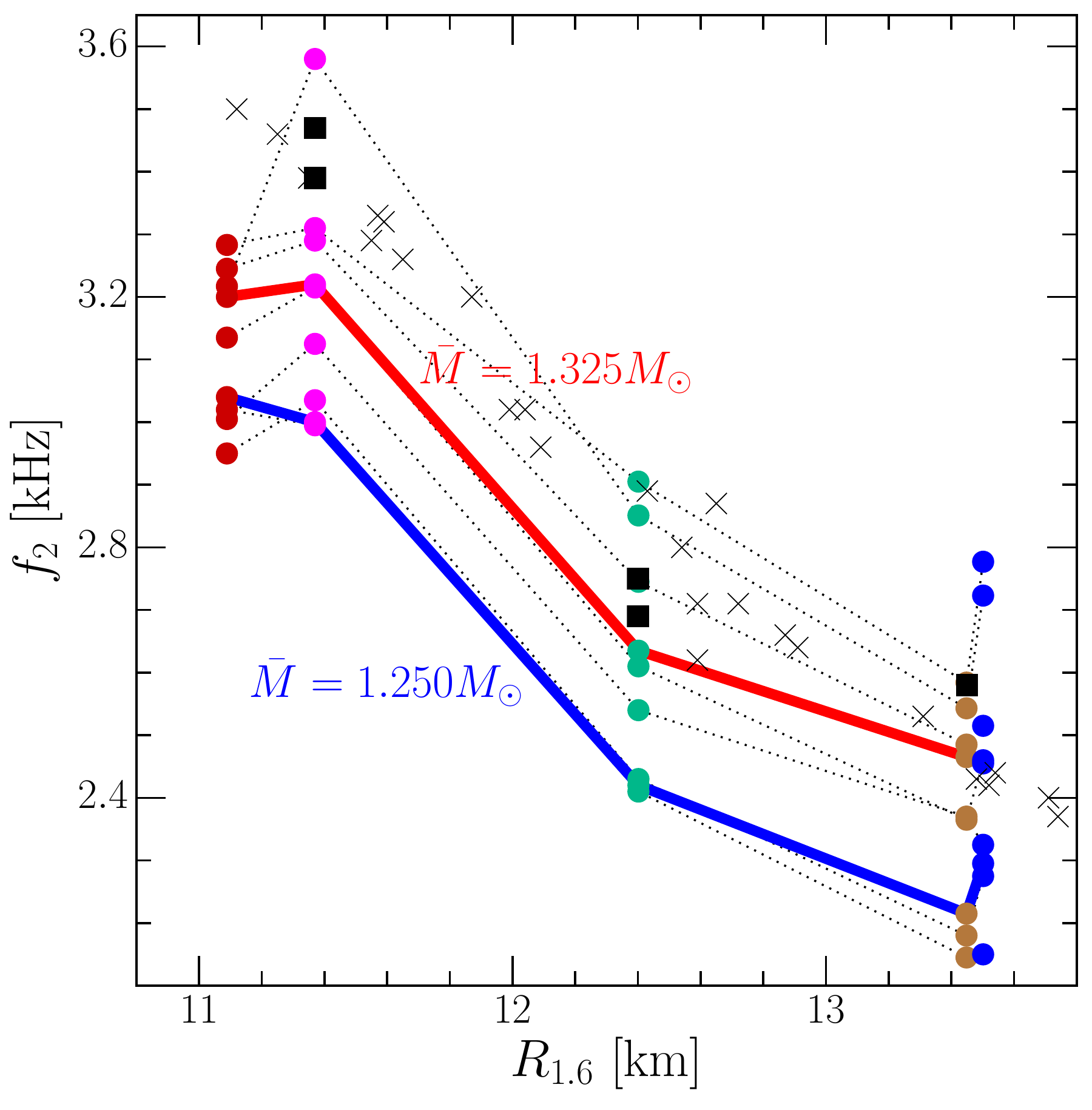}
\caption{Values of the $f_2$ frequencies shown in terms of $R_{1.6}$, the
  radius of a nonrotating neutron star with $\bar{M}=1.6\,M_\odot$. The
  various filled circles refer to the $f_2$ frequencies at masses
  $\bar{M}/M_{\odot}=1.200, 1.225, 1.250, 1.275, 1.300, 1.325, 1.350,
  1.375$, and $1.400$; these sequences are connected with dotted lines,
  while the thick blue and red lines refer to our reference sequences
  having mass $\bar{M}/M_{\odot}=1.250$ and $\bar{M}/M_{\odot}=1.325$,
  respectively. Indicated with black crosses is the data reported in
  Refs.~\cite{Bauswein2012, Clark2016} for equal-mass binaries with
  $\bar{M}=1.35\,M_{\odot}$, while filled squares refer to the data of
  Ref. \cite{Bernuzzi2015a} with $\bar{M} = \tfrac{1}{2}(1.35+1.35)
  M_{\odot}$ and $\bar{M} = \tfrac{1}{2}(1.25+1.45) M_{\odot}$. Note the
  large scatter across different masses.}
\label{fig:f2_vs_R16}
\end{center}
\end{figure}

We also performed a set of Monte-Carlo simulations where the masses were
allowed to span a wider range, namely, $[1.0, 2.0]\,M_\odot$, for all
four EOSs and the two mass distributions. Since we do not yet have
numerical-relativity data on the $f_2$ and compactness for masses beyond
$1.4~M_\odot$, we extrapolated the quasi-universal relations to infer
them. In these studies, there is no appreciable change in the radius
errors for the GNH3 and H4 EOSs, but for ALF2 and SLy the radius errors
rise to $7.4\%$ (Gaussian) and $5.4\%$ (uniform) and $12.0\%$ (Gaussian)
and $22.0\%$ (uniform), respectively, at 90\% confidence level. 
These large errors are not
surprising since we expect waveforms from softer stars to damp faster and
have the $f_2$ peak at frequencies where the aLIGO sensitivity is worse.
They warn us that our error estimates may change once
numerical-relativity simulations for heavier stars become available.

Obviously, the Monte-Carlo results are more trustworthy than the Fisher
results, but the latter are useful in providing a partial understanding
of the former. It is also true that the size of the errors depends on
the measurement method employed. For instance, a coherent way of
combining the EOS effects on full signal, from inspiral to postmerger,
will influence those errors and impact EOS discernability. 

A concluding remark will be dedicated to whether or not it is possible to
find different and tighter correlations between the spectral properties
of the postmerger signal and the radius at a given fixed mass. Results
presented in Ref.~\cite{Clark2016}, in fact, suggested that it is
possible to find a quadratic quasi-universal relation between the $f_2$
frequency and the radius of nonrotating neutron stars with a fixed mass of
$1.6\,M_{\odot}$ [see Eq. (2) in Ref.~\cite{Clark2016}, which the fit
  reported in the right panel of Fig. 4 in Ref.~\cite{Clark2016}].

When considering the results of our simulations, but also those of other
groups, we are not able to confirm this behaviour; rather, we find that
the scattering in the frequencies is quite large. This is shown in
Fig.~\ref{fig:f2_vs_R16}, which reports the values of the $f_2$
frequencies for the various binaries considered in this work using the
same colorcode adopted in all other figures, together with other data
collected from the literature, with black crosses referring to the data
reported in Refs.~\cite{Bauswein2012, Clark2016} for equal-mass binaries
with $\bar{M}=1.35\,M_{\odot}$, while filled squares refer to
the data of Ref. \cite{Bernuzzi2015a} with $\bar{M}=1.35\,M_{\odot}$ but
not necessarily for equal-mass binaries (\ie
$\bar{M}=\tfrac{1}{2}(1.35+1.35)M_{\odot}$ and
$\bar{M}=\tfrac{1}{2}(1.25+1.45)M_{\odot}$). To help the comparison with
Refs.~\cite{Bauswein2012, Clark2016} we present the results for sequences
with constant masses of $\bar{M}/M_{\odot}=1.200, 1.225, 1.250, 1.275,
1.300, 1.325, 1.350, 1.375$, and $1.400$. The different sequences are
connected with dotted lines, while the thick blue and red lines refer to
our reference sequences having mass $\bar{M}/M_{\odot}=1.250$ and
$\bar{M}/M_{\odot}=1.325$, respectively.

As is apparent from Fig.~\ref{fig:f2_vs_R16}, we cannot confirm that
the quadratic fitting suggested by Eq.~(2) of Ref.~\cite{Clark2016}
is a good representation of the data, which instead show a
rather large scatter. It is presently unclear what the origin of this
discrepancy is, but our results, combined with those of other groups using
fully general-relativistic numerical codes, warn against making use of
quasi-universal relations for sequences of fixed gravitational
mass. Specifically, owing to the mass-related spread of the curves in
Fig.~\ref{fig:f2_vs_R16} it is not obvious how for the population of
$100$ BNSs one might estimate $R_{1.6}$ more accurately than the average
radius of that population from the methods presented in this letter. 
However, for the nearby
binary at $30\,{\rm Mpc}$ discussed above, since the BNS masses can be
estimated to a high accuracy, it is possible to narrow that
spread; this allows the determination of $R_{1.6}$ for that case with an
accuracy that rivals the estimation of the average radius deduced
above.


\end{document}